\documentstyle[12pt]{article}
\begin{document}

\overfullrule 0 mm
\language 0

\rightline{ \it{In Memory of my Beloved Mama,}}
\rightline{ \it{Vlasova Alexandra V.,}}
\rightline{ \it{1922-2000}}
\vskip 0.5 cm

\centerline { \bf{ CHARGED BROWN PARTICLE: }}
\centerline { \bf{THE MORE RETARDATION IS}}
\centerline { \bf{ - THE LOWER IS THE EFFECTIVE TEMPERATURE }}
\vskip 0.5 cm \centerline {\bf{ Alexander A.  Vlasov}}
\vskip 0.3 cm \centerline {{  High Energy and Quantum Theory}}
\centerline {{  Department of Physics}} \centerline {{ Moscow State
University}} \centerline {{  Moscow, 119899}} \centerline {{
Russia}} \vskip 0.3 cm
 {\it  The Brownian motion of a charged particle with finite size 
(described by Sommerfeld model) is considered. It is found out that due to 
radiation reaction:  (1) the effective temperature of such particle is 
lower, and (2) the acceleration of the average velocity is   smaller, then 
that for classical Brown particle without electric charge.}

03.50.De
\vskip 0.3 cm

 Sommerfeld particle  [1] is the model of a charged particle of finite 
size - sphere with uniform surface charge $Q$ and mechanical mass $m$.  In 
nonrelativistic approximation such sphere obeys the equation (see  [2] 
and works, cited there): 
 $$m\dot{\vec v} =\vec F_{ext}+  \eta\left[\vec 
v(t-2a/c) - \vec v(t)\right] \eqno(1)$$ here $a$ - radius of the sphere, 
$\eta= {Q^2 \over 3  c a^2},\ \ \vec v= d \vec R /dt,\ \ \vec R$ - 
 coordinate of the center of the shell, $\vec F_{ext}$ - some external 
force. The term in RHS of eq.(1), proportional to $\eta$,  describes the 
effect of radiation reaction (effect of retardation).

This model is a good tool to consider effects of radiation reaction,  free 
of problems of classical point-like Lorentz-Dirac description (see, for 
ex.[3]).
  \vskip 0.3 cm {\bf {A.}} \vskip 0.3 cm
  In this paper we consider      Sommerfeld particle in the role of the 
Brown particle, i.e. particle (with  radiation reaction), 
moving in a stochastic path under action of some external stochastic force 
$\vec F_{stoch}$.

 For simplicity we take that:

 (1) the viscosity of the surrounding medium is zero;

 (2) the statistical average of $\vec F_{stoch}$ - $\vec F_{0}$ - 
is non zero and constant in time $t$;

 (3) particle moves in one dimension.

Under these assumptions the Langevin equation for Sommerfeld particle 
 takes the form

  $$m\dot{ v} = F_{stoch}+  \eta\left[ v(t-2a/c) -  
v(t)\right] \eqno(2)$$

or for statistical average value $<v>$  of $v$

  $$m\dot{ <v>} = <F_{stoch}>+  \eta\left[ <v>(t-2a/c) -  
<v>(t)\right] \eqno(3)$$

For dimensionless variables $y=<v>/M,\ \ \ \ x =ct/M$ ($M$
-scale factor) equation (3) takes the form 
$$\dot y
=f+\gamma \cdot\left[ y(x-\delta) - y(x)\right] \eqno (4)$$ 
here
$$\gamma= { Q^2 M\over 3 a^2 m c^2},\ \ f={F_{0}M\over mc^2},\ \
\delta={2a\over M}$$
Classical analog of equation (4) for Brown point particle one can get 
taking $\gamma=0$ in (4).

Equation (4) for $f=const$  has the exact solution
$$ y =y_0 +k \cdot x \eqno(5)$$
with $k= f/(1+\gamma \delta)$ and initial velocity $y_0$.

Following the theory of Brownian motion   ( there are many textbooks on 
Brownian motion, see, for ex., [4]), the dispersion $D=<(v-<v>)^2>$ for 
surrounding medium without viscosity can be function of time: $D=D(t)$. 
The form of $D(t)$ strongly depends on the form of correlation 
function of stochastic force $F_{stoch}$ and its concrete realization in 
computer program ( if correlation function is compact enough, then $D(t)$  
for "not very large moments of time $t$" is proportional to time: $D(t) 
\sim t$ - Einstein formula for Brownian motion).  If the dependence 
$D=D(t)$ is known, one can try to find the form of correlation function, 
but this is not our goal.

Do not going into details, we can say that time average of $D$ with 
respect to the whole time of "observation" $T$ must be proportional (in 
dimensionless variables) to the effective temperature $\theta$ and inverse 
to the mass of particle:
$$ \bar D \sim  {\theta \over mc^2}. \eqno (6)$$

Solution (5) describes motion of charged particle with constant 
acceleration. This result seems to be  unusual.

Indeed, following classical electrodynamics, particle with acceleration 
must radiate. Then one could expect that radiation, due to energy loss, 
leads to radiation damping of particle motion. 

Instead, we see that after 
statistical averaging, radiation reaction in Sommerfeld form only changes 
the value of the effective force, acting on particle, making it smaller.

One can interpret this in the following way. Radiation reaction $F_{rad}$ 
in Sommerfeld model is $$F_{rad}=\gamma \cdot\left[ y(x-\delta) - 
y(x)\right] $$ on trajectory (5) it is nonzero and equals to $-\gamma 
\delta k $ . If $\delta$ - is small we can expand the force $F_{rad}$  in 
powers of $\delta$:  $$F_{rad} = -k\delta \dot y+k/2(\delta)^2\ddot y 
+...\eqno(7)$$

First term in (7)  is the effective electromagnetic mass of the 
particle, multiplied by acceleration with sign minus. On trajectory (5) 
this term equals to $-\gamma \delta k $. Second term in (7) is the 
radiation force in classical Lorentz form - on trajectory (7) it is zero.
Thus radiation effects after statistical averaging lead only to change in 
effective particle mass (mechanic + electromagnetic) - it becomes 
greater and the particle becomes more "inertial" (more "retarded").

It looks like the energy of self-electromagnetic field of a charged 
particle does not radiate away, the "electromagnetic fur-coat" does not 
get thin, thus particle becomes "heavier" (in comparison with particle of 
zero charge).

This effect can also make the dispersion (6), i.e. the effective Brownian 
temperature smaller.

\vskip 0.3 cm
{\bf {B.}}
\vskip 0.3 cm
 
 Dispersion (6), also as the particle motion, we investigated  
numerically. The particle mass $m$ and size $a$ we take close to that of 
classical electron, this yields $\gamma \delta =1$ in (4).

The stochastic force $F_{stoch}$ we extract step by step from the known 
procedure (see, for ex.,[5]):
 $$\phi_{n+1}=\left \{ K \cdot \phi_{n} \right \};$$
 here brackets $\{...\}$ denotes fractional part of $...$,
 
  and
$$ (F_{stoch})_n\cdot (M/mc^2)=10^{+3} \cdot (\phi_{n} -\phi_{0})$$

 with $K=100000/3$ and $\phi_{0}=0.5007645$.

 The results of numerical calculations can be summed in the following way:

(1) Sommerfeld particle in the role of the Brown particle has the 
effective temperature $\theta_{S}$ lower, than that for classical Brown 
particle $\theta_{B}$:  $\theta_{S}<\theta_{B}$.

(2) The more is the retardation (i.e. the greater is $\gamma$ in (4) ), 
the greater is the difference between $\theta_{B}$ and $\theta_{S}$.

(3) The acceleration $k$ of the average velocity of Sommerfeld particle is 
smaller than that of classical Brown particle without electric charge. 

These results are illustrated in  Fig. 1, where $y_0=0.1;\ \ \gamma 
=2000.0;\ \ \delta =2a/M= 1/2000.0 $; observation "time" is $10.0005$, 
with respect to it the average value of stochastic force is $\ \ 
f=4.881996 \cdot 10^{-4}$  and the average dispersions are ${\bar 
D}_{S}=4.4359\cdot 10^{-8},\ \ \ \ {\bar D}_{B}=4.8496\cdot 10^{-8} $, 
i.e. $ \theta_{S}< \theta_{B}$.

The upper curve - is the path of classical Brown particle (horizontal 
axis - is "time" $x$), the lower - is the path of Sommerfeld particle in 
the role of Brown particle.
 
The acceleration of the average velocities for these particles differs by 
multiplier $2$, as the consequence of the exact solution (5) ( for Brown 
particle we have $\gamma=0$, and $k=f$,  for Sommerfeld - $\gamma \delta 
=1$ and $k=f/2$).

These results confirm numerically our considerations, made before.
\eject
 \vskip 2 cm \centerline {\bf{REFERENCES}}

  \begin{enumerate}

\item A.Sommerfeld, Gottingen Nachrichten, 29 (1904), 363 (1904), 201
  (1905).
\item L.Page, Phys.Rev., 11, 377 (1918).
 T.Erber, Fortschr. Phys., 9, 343 (1961).
 P.Pearle in "Electromagnetism",ed. D.Tepliz, (Plenum, N.Y.,
1982), p.211.
 A.Yaghjian, "Relativistic Dynamics of a Charged Sphere".
  Lecture Notes in Physics, 11 (Springer-Verlag, Berlin, 1992).
F.Rohrlich, Phys.Rev.D, 60, 084017 (1999).
\item Alexander A.Vlasov, physics/9905050, physics/9911059, 
physics/0004026.

\item I.A.Kvasnikov, "Thermodynamics and statistical physics. Part 
2".  Moscow, Moscow State University, 1987.

\item R.Z.Sagdeev, G.M.Zaslavsky, "Introduction to nonlinear 
 physics".  Moscow, Nauka, 1988.  \end{enumerate} \eject

\newcount\numpoint                     
\newcount\numpointo                    
\numpoint=1 \numpointo=1               
\def\emmoveto#1#2{\offinterlineskip   
\hbox to 0 true cm{\vbox to 0          
true cm{\vskip - #2 true mm             
\hskip #1 true mm \special{em:point    
\the\numpoint}\vss}\hss}             
\numpointo=\numpoint                   
\global\advance \numpoint by 1}       
\def\emlineto#1#2{\offinterlineskip   
\hbox to 0 true cm{\vbox to 0          
true cm{\vskip - #2 true mm             
\hskip #1 true mm \special{em:point    
\the\numpoint}\vss}\hss}             
\special{em:line                        
\the\numpointo,\the\numpoint}        
\numpointo=\numpoint                   
\global\advance \numpoint by 1}       
\def\emshow#1#2#3{\offinterlineskip   
\hbox to 0 true cm{\vbox to 0          
true cm{\vskip - #2 true mm             
\hskip #1 true mm \vbox to 0           
true cm{\vss\hbox{#3\hss              
}}\vss}\hss}}                          
\special{em:linewidth 0.8pt}            

\vrule width 0 mm height                0 mm depth 90.000 true mm                

\special{em:linewidth 0.8pt} 
\emmoveto{130.000}{10.000} 
\emlineto{12.000}{10.000} 
\emlineto{12.000}{80.000} 
\emmoveto{130.000}{10.000} 
\emlineto{130.000}{80.000} 
\emlineto{12.000}{80.000} 
\emlineto{12.000}{10.000} 
\emlineto{130.000}{10.000} 
\special{em:linewidth 0.4pt} 
\emmoveto{12.000}{24.000} 
\emlineto{130.000}{24.000} 
\emmoveto{12.000}{38.000} 
\emlineto{130.000}{38.000} 
\emmoveto{12.000}{52.000} 
\emlineto{130.000}{52.000} 
\emmoveto{12.000}{66.000} 
\emlineto{130.000}{66.000} 
\emmoveto{23.800}{10.000} 
\emlineto{23.800}{80.000} 
\emmoveto{35.600}{10.000} 
\emlineto{35.600}{80.000} 
\emmoveto{47.400}{10.000} 
\emlineto{47.400}{80.000} 
\emmoveto{59.200}{10.000} 
\emlineto{59.200}{80.000} 
\emmoveto{71.000}{10.000} 
\emlineto{71.000}{80.000} 
\emmoveto{82.800}{10.000} 
\emlineto{82.800}{80.000} 
\emmoveto{94.600}{10.000} 
\emlineto{94.600}{80.000} 
\emmoveto{106.400}{10.000} 
\emlineto{106.400}{80.000} 
\emmoveto{118.200}{10.000} 
\emlineto{118.200}{80.000} 
\special{em:linewidth 0.8pt} 
\emmoveto{12.000}{14.073} 
\emlineto{12.596}{14.245} 
\emmoveto{12.596}{14.235} 
\emlineto{13.186}{14.406} 
\emmoveto{13.186}{14.396} 
\emlineto{13.776}{14.567} 
\emmoveto{13.776}{14.557} 
\emlineto{14.366}{14.728} 
\emmoveto{14.366}{14.718} 
\emlineto{14.956}{14.889} 
\emmoveto{14.956}{14.879} 
\emlineto{15.546}{15.051} 
\emmoveto{15.546}{15.041} 
\emlineto{16.136}{15.212} 
\emmoveto{16.136}{15.202} 
\emlineto{16.726}{15.373} 
\emmoveto{16.726}{15.363} 
\emlineto{17.316}{15.533} 
\emmoveto{17.316}{15.523} 
\emlineto{17.906}{15.694} 
\emmoveto{17.906}{15.684} 
\emlineto{18.496}{15.855} 
\emmoveto{18.496}{15.845} 
\emlineto{19.086}{16.016} 
\emmoveto{19.086}{16.006} 
\emlineto{19.676}{16.177} 
\emmoveto{19.676}{16.167} 
\emlineto{20.265}{16.338} 
\emmoveto{20.265}{16.328} 
\emlineto{20.855}{16.499} 
\emmoveto{20.855}{16.489} 
\emlineto{21.445}{16.659} 
\emmoveto{21.445}{16.649} 
\emlineto{22.035}{16.820} 
\emmoveto{22.035}{16.810} 
\emlineto{22.625}{16.981} 
\emmoveto{22.625}{16.971} 
\emlineto{23.215}{17.142} 
\emmoveto{23.215}{17.132} 
\emlineto{23.805}{17.303} 
\emmoveto{23.805}{17.293} 
\emlineto{24.395}{17.464} 
\emmoveto{24.395}{17.454} 
\emlineto{24.985}{17.625} 
\emmoveto{24.985}{17.615} 
\emlineto{25.575}{17.785} 
\emmoveto{25.575}{17.775} 
\emlineto{26.165}{17.946} 
\emmoveto{26.165}{17.936} 
\emlineto{26.755}{18.107} 
\emmoveto{26.755}{18.097} 
\emlineto{27.345}{18.269} 
\emmoveto{27.345}{18.259} 
\emlineto{27.935}{18.430} 
\emmoveto{27.935}{18.420} 
\emlineto{28.525}{18.591} 
\emmoveto{28.525}{18.581} 
\emlineto{29.115}{18.752} 
\emmoveto{29.115}{18.742} 
\emlineto{29.705}{18.913} 
\emmoveto{29.705}{18.903} 
\emlineto{30.295}{19.074} 
\emmoveto{30.295}{19.064} 
\emlineto{30.885}{19.235} 
\emmoveto{30.885}{19.225} 
\emlineto{31.475}{19.395} 
\emmoveto{31.475}{19.385} 
\emlineto{32.065}{19.556} 
\emmoveto{32.065}{19.546} 
\emlineto{32.655}{19.717} 
\emmoveto{32.655}{19.707} 
\emlineto{33.245}{19.878} 
\emmoveto{33.245}{19.868} 
\emlineto{33.835}{20.039} 
\emmoveto{33.835}{20.029} 
\emlineto{34.425}{20.200} 
\emmoveto{34.425}{20.190} 
\emlineto{35.015}{20.361} 
\emmoveto{35.015}{20.351} 
\emlineto{35.605}{20.521} 
\emmoveto{35.605}{20.511} 
\emlineto{36.195}{20.682} 
\emmoveto{36.195}{20.672} 
\emlineto{36.785}{20.843} 
\emmoveto{36.785}{20.833} 
\emlineto{37.375}{21.004} 
\emmoveto{37.375}{20.994} 
\emlineto{37.965}{21.165} 
\emmoveto{37.965}{21.155} 
\emlineto{38.555}{21.326} 
\emmoveto{38.555}{21.316} 
\emlineto{39.145}{21.488} 
\emmoveto{39.145}{21.478} 
\emlineto{39.735}{21.649} 
\emmoveto{39.735}{21.639} 
\emlineto{40.324}{21.810} 
\emmoveto{40.324}{21.800} 
\emlineto{40.914}{21.970} 
\emmoveto{40.914}{21.960} 
\emlineto{41.504}{22.131} 
\emmoveto{41.504}{22.121} 
\emlineto{42.094}{22.292} 
\emmoveto{42.094}{22.282} 
\emlineto{42.684}{22.453} 
\emmoveto{42.684}{22.443} 
\emlineto{43.274}{22.614} 
\emmoveto{43.274}{22.604} 
\emlineto{43.864}{22.775} 
\emmoveto{43.864}{22.765} 
\emlineto{44.454}{22.936} 
\emmoveto{44.454}{22.926} 
\emlineto{45.044}{23.097} 
\emmoveto{45.044}{23.087} 
\emlineto{45.634}{23.257} 
\emmoveto{45.634}{23.247} 
\emlineto{46.224}{23.418} 
\emmoveto{46.224}{23.408} 
\emlineto{46.814}{23.579} 
\emmoveto{46.814}{23.569} 
\emlineto{47.404}{23.740} 
\emmoveto{47.404}{23.730} 
\emlineto{47.994}{23.901} 
\emmoveto{47.994}{23.891} 
\emlineto{48.584}{24.062} 
\emmoveto{48.584}{24.052} 
\emlineto{49.174}{24.223} 
\emmoveto{49.174}{24.213} 
\emlineto{49.764}{24.383} 
\emmoveto{49.764}{24.373} 
\emlineto{50.354}{24.544} 
\emmoveto{50.354}{24.534} 
\emlineto{50.944}{24.706} 
\emmoveto{50.944}{24.696} 
\emlineto{51.534}{24.867} 
\emmoveto{51.534}{24.857} 
\emlineto{52.124}{25.028} 
\emmoveto{52.124}{25.018} 
\emlineto{52.714}{25.189} 
\emmoveto{52.714}{25.179} 
\emlineto{53.304}{25.350} 
\emmoveto{53.304}{25.340} 
\emlineto{53.894}{25.511} 
\emmoveto{53.894}{25.501} 
\emlineto{54.484}{25.672} 
\emmoveto{54.484}{25.662} 
\emlineto{55.074}{25.832} 
\emmoveto{55.074}{25.822} 
\emlineto{55.664}{25.993} 
\emmoveto{55.664}{25.983} 
\emlineto{56.254}{26.154} 
\emmoveto{56.254}{26.144} 
\emlineto{56.844}{26.315} 
\emmoveto{56.844}{26.305} 
\emlineto{57.434}{26.476} 
\emmoveto{57.434}{26.466} 
\emlineto{58.024}{26.637} 
\emmoveto{58.024}{26.627} 
\emlineto{58.614}{26.798} 
\emmoveto{58.614}{26.788} 
\emlineto{59.204}{26.959} 
\emmoveto{59.204}{26.949} 
\emlineto{59.794}{27.119} 
\emmoveto{59.794}{27.109} 
\emlineto{60.383}{27.280} 
\emmoveto{60.383}{27.270} 
\emlineto{60.973}{27.441} 
\emmoveto{60.973}{27.431} 
\emlineto{61.563}{27.602} 
\emmoveto{61.563}{27.592} 
\emlineto{62.153}{27.763} 
\emmoveto{62.153}{27.753} 
\emlineto{62.743}{27.925} 
\emmoveto{62.743}{27.915} 
\emlineto{63.333}{28.086} 
\emmoveto{63.333}{28.076} 
\emlineto{63.923}{28.247} 
\emmoveto{63.923}{28.237} 
\emlineto{64.513}{28.408} 
\emmoveto{64.513}{28.398} 
\emlineto{65.103}{28.568} 
\emmoveto{65.103}{28.558} 
\emlineto{65.693}{28.729} 
\emmoveto{65.693}{28.719} 
\emlineto{66.283}{28.890} 
\emmoveto{66.283}{28.880} 
\emlineto{66.873}{29.051} 
\emmoveto{66.873}{29.041} 
\emlineto{67.463}{29.212} 
\emmoveto{67.463}{29.202} 
\emlineto{68.053}{29.373} 
\emmoveto{68.053}{29.363} 
\emlineto{68.643}{29.534} 
\emmoveto{68.643}{29.524} 
\emlineto{69.233}{29.694} 
\emmoveto{69.233}{29.684} 
\emlineto{69.823}{29.855} 
\emmoveto{69.823}{29.845} 
\emlineto{70.413}{30.016} 
\emmoveto{70.413}{30.006} 
\emlineto{71.003}{30.177} 
\emmoveto{71.003}{30.167} 
\emlineto{71.593}{30.338} 
\emmoveto{71.593}{30.328} 
\emlineto{72.183}{30.499} 
\emmoveto{72.183}{30.489} 
\emlineto{72.773}{30.660} 
\emmoveto{72.773}{30.650} 
\emlineto{73.363}{30.821} 
\emmoveto{73.363}{30.811} 
\emlineto{73.953}{30.981} 
\emmoveto{73.953}{30.971} 
\emlineto{74.543}{31.144} 
\emmoveto{74.543}{31.134} 
\emlineto{75.133}{31.304} 
\emmoveto{75.133}{31.294} 
\emlineto{75.723}{31.465} 
\emmoveto{75.723}{31.455} 
\emlineto{76.313}{31.626} 
\emmoveto{76.313}{31.616} 
\emlineto{76.903}{31.787} 
\emmoveto{76.903}{31.777} 
\emlineto{77.493}{31.948} 
\emmoveto{77.493}{31.938} 
\emlineto{78.083}{32.109} 
\emmoveto{78.083}{32.099} 
\emlineto{78.673}{32.270} 
\emmoveto{78.673}{32.260} 
\emlineto{79.263}{32.430} 
\emmoveto{79.263}{32.420} 
\emlineto{79.853}{32.591} 
\emmoveto{79.853}{32.581} 
\emlineto{80.442}{32.752} 
\emmoveto{80.442}{32.742} 
\emlineto{81.032}{32.913} 
\emmoveto{81.032}{32.903} 
\emlineto{81.622}{33.074} 
\emmoveto{81.622}{33.064} 
\emlineto{82.212}{33.235} 
\emmoveto{82.212}{33.225} 
\emlineto{82.802}{33.396} 
\emmoveto{82.802}{33.386} 
\emlineto{83.392}{33.556} 
\emmoveto{83.392}{33.546} 
\emlineto{83.982}{33.717} 
\emmoveto{83.982}{33.707} 
\emlineto{84.572}{33.878} 
\emmoveto{84.572}{33.868} 
\emlineto{85.162}{34.039} 
\emmoveto{85.162}{34.029} 
\emlineto{85.752}{34.200} 
\emmoveto{85.752}{34.190} 
\emlineto{86.342}{34.362} 
\emmoveto{86.342}{34.352} 
\emlineto{86.932}{34.523} 
\emmoveto{86.932}{34.513} 
\emlineto{87.522}{34.684} 
\emmoveto{87.522}{34.674} 
\emlineto{88.112}{34.845} 
\emmoveto{88.112}{34.835} 
\emlineto{88.702}{35.006} 
\emmoveto{88.702}{34.996} 
\emlineto{89.292}{35.166} 
\emmoveto{89.292}{35.156} 
\emlineto{89.882}{35.327} 
\emmoveto{89.882}{35.317} 
\emlineto{90.472}{35.488} 
\emmoveto{90.472}{35.478} 
\emlineto{91.062}{35.649} 
\emmoveto{91.062}{35.639} 
\emlineto{91.652}{35.810} 
\emmoveto{91.652}{35.800} 
\emlineto{92.242}{35.971} 
\emmoveto{92.242}{35.961} 
\emlineto{92.832}{36.132} 
\emmoveto{92.832}{36.122} 
\emlineto{93.422}{36.292} 
\emmoveto{93.422}{36.282} 
\emlineto{94.012}{36.453} 
\emmoveto{94.012}{36.443} 
\emlineto{94.602}{36.614} 
\emmoveto{94.602}{36.604} 
\emlineto{95.192}{36.775} 
\emmoveto{95.192}{36.765} 
\emlineto{95.782}{36.936} 
\emmoveto{95.782}{36.926} 
\emlineto{96.372}{37.097} 
\emmoveto{96.372}{37.087} 
\emlineto{96.962}{37.258} 
\emmoveto{96.962}{37.248} 
\emlineto{97.552}{37.419} 
\emmoveto{97.552}{37.409} 
\emlineto{98.142}{37.581} 
\emmoveto{98.142}{37.571} 
\emlineto{98.732}{37.742} 
\emmoveto{98.732}{37.732} 
\emlineto{99.322}{37.902} 
\emmoveto{99.322}{37.892} 
\emlineto{99.912}{38.063} 
\emmoveto{99.912}{38.053} 
\emlineto{100.501}{38.224} 
\emmoveto{100.501}{38.214} 
\emlineto{101.091}{38.385} 
\emmoveto{101.091}{38.375} 
\emlineto{101.681}{38.546} 
\emmoveto{101.681}{38.536} 
\emlineto{102.271}{38.707} 
\emmoveto{102.271}{38.697} 
\emlineto{102.861}{38.868} 
\emmoveto{102.861}{38.858} 
\emlineto{103.451}{39.028} 
\emmoveto{103.451}{39.018} 
\emlineto{104.041}{39.189} 
\emmoveto{104.041}{39.179} 
\emlineto{104.631}{39.350} 
\emmoveto{104.631}{39.340} 
\emlineto{105.221}{39.511} 
\emmoveto{105.221}{39.501} 
\emlineto{105.811}{39.672} 
\emmoveto{105.811}{39.662} 
\emlineto{106.401}{39.833} 
\emmoveto{106.401}{39.823} 
\emlineto{106.991}{39.994} 
\emmoveto{106.991}{39.984} 
\emlineto{107.581}{40.154} 
\emmoveto{107.581}{40.144} 
\emlineto{108.171}{40.315} 
\emmoveto{108.171}{40.305} 
\emlineto{108.761}{40.476} 
\emmoveto{108.761}{40.466} 
\emlineto{109.351}{40.637} 
\emmoveto{109.351}{40.627} 
\emlineto{109.941}{40.799} 
\emmoveto{109.941}{40.789} 
\emlineto{110.531}{40.960} 
\emmoveto{110.531}{40.950} 
\emlineto{111.121}{41.121} 
\emmoveto{111.121}{41.111} 
\emlineto{111.711}{41.282} 
\emmoveto{111.711}{41.272} 
\emlineto{112.301}{41.443} 
\emmoveto{112.301}{41.433} 
\emlineto{112.891}{41.604} 
\emmoveto{112.891}{41.594} 
\emlineto{113.481}{41.764} 
\emmoveto{113.481}{41.754} 
\emlineto{114.071}{41.925} 
\emmoveto{114.071}{41.915} 
\emlineto{114.661}{42.086} 
\emmoveto{114.661}{42.076} 
\emlineto{115.251}{42.247} 
\emmoveto{115.251}{42.237} 
\emlineto{115.841}{42.408} 
\emmoveto{115.841}{42.398} 
\emlineto{116.431}{42.569} 
\emmoveto{116.431}{42.559} 
\emlineto{117.021}{42.730} 
\emmoveto{117.021}{42.720} 
\emlineto{117.611}{42.890} 
\emmoveto{117.611}{42.880} 
\emlineto{118.201}{43.051} 
\emmoveto{118.201}{43.041} 
\emlineto{118.791}{43.212} 
\emmoveto{118.791}{43.202} 
\emlineto{119.381}{43.373} 
\emmoveto{119.381}{43.363} 
\emlineto{119.971}{43.534} 
\emmoveto{119.971}{43.524} 
\emlineto{120.560}{43.695} 
\emmoveto{120.560}{43.685} 
\emlineto{121.150}{43.856} 
\emmoveto{121.150}{43.846} 
\emlineto{121.740}{44.018} 
\emmoveto{121.740}{44.008} 
\emlineto{122.330}{44.179} 
\emmoveto{122.330}{44.169} 
\emlineto{122.920}{44.340} 
\emmoveto{122.920}{44.330} 
\emlineto{123.510}{44.500} 
\emmoveto{123.510}{44.490} 
\emlineto{124.100}{44.661} 
\emmoveto{124.100}{44.651} 
\emlineto{124.690}{44.822} 
\emmoveto{124.690}{44.812} 
\emlineto{125.280}{44.983} 
\emmoveto{125.280}{44.973} 
\emlineto{125.870}{45.144} 
\emmoveto{125.870}{45.134} 
\emlineto{126.460}{45.305} 
\emmoveto{126.460}{45.295} 
\emlineto{127.050}{45.466} 
\emmoveto{127.050}{45.456} 
\emlineto{127.640}{45.626} 
\emmoveto{127.640}{45.616} 
\emlineto{128.230}{45.787} 
\emmoveto{128.230}{45.777} 
\emlineto{128.820}{45.948} 
\emmoveto{128.820}{45.938} 
\emlineto{129.410}{46.109} 
\emshow{24.980}{25.400}{} 
\emmoveto{12.000}{14.073} 
\emlineto{12.596}{10.295} 
\emlineto{12.596}{10.305} 
\emmoveto{12.596}{10.295} 
\emlineto{13.186}{13.584} 
\emlineto{13.186}{13.594} 
\emmoveto{13.186}{13.584} 
\emlineto{13.776}{20.651} 
\emlineto{13.776}{20.661} 
\emmoveto{13.776}{20.651} 
\emlineto{14.366}{15.547} 
\emlineto{14.366}{15.557} 
\emmoveto{14.366}{15.547} 
\emlineto{14.956}{19.591} 
\emlineto{14.956}{19.601} 
\emmoveto{14.956}{19.591} 
\emlineto{15.546}{11.619} 
\emlineto{15.546}{11.629} 
\emmoveto{15.546}{11.619} 
\emlineto{16.136}{17.203} 
\emlineto{16.136}{17.213} 
\emmoveto{16.136}{17.203} 
\emlineto{16.726}{18.394} 
\emlineto{16.726}{18.404} 
\emmoveto{16.726}{18.394} 
\emlineto{17.316}{10.000} 
\emlineto{17.316}{10.010} 
\emmoveto{17.316}{10.000} 
\emlineto{17.906}{14.169} 
\emlineto{17.906}{14.179} 
\emmoveto{17.906}{14.169} 
\emlineto{18.496}{14.813} 
\emlineto{18.496}{14.823} 
\emmoveto{18.496}{14.813} 
\emlineto{19.086}{17.828} 
\emlineto{19.086}{17.838} 
\emmoveto{19.086}{17.828} 
\emlineto{19.676}{19.773} 
\emlineto{19.676}{19.783} 
\emmoveto{19.676}{19.773} 
\emlineto{20.265}{15.410} 
\emlineto{20.265}{15.420} 
\emmoveto{20.265}{15.410} 
\emlineto{20.855}{16.520} 
\emlineto{20.855}{16.530} 
\emmoveto{20.855}{16.520} 
\emlineto{21.445}{15.684} 
\emlineto{21.445}{15.694} 
\emmoveto{21.445}{15.684} 
\emlineto{22.035}{17.013} 
\emlineto{22.035}{17.023} 
\emmoveto{22.035}{17.013} 
\emlineto{22.625}{21.665} 
\emlineto{22.625}{21.675} 
\emmoveto{22.625}{21.665} 
\emlineto{23.215}{20.225} 
\emlineto{23.215}{20.235} 
\emmoveto{23.215}{20.225} 
\emlineto{23.805}{19.624} 
\emlineto{23.805}{19.634} 
\emmoveto{23.805}{19.624} 
\emlineto{24.395}{18.908} 
\emlineto{24.395}{18.918} 
\emmoveto{24.395}{18.908} 
\emlineto{24.985}{17.009} 
\emlineto{24.985}{17.019} 
\emmoveto{24.985}{17.009} 
\emlineto{25.575}{20.203} 
\emlineto{25.575}{20.213} 
\emmoveto{25.575}{20.203} 
\emlineto{26.165}{14.243} 
\emlineto{26.165}{14.253} 
\emmoveto{26.165}{14.243} 
\emlineto{26.755}{16.682} 
\emlineto{26.755}{16.692} 
\emmoveto{26.755}{16.682} 
\emlineto{27.345}{16.959} 
\emlineto{27.345}{16.969} 
\emmoveto{27.345}{16.959} 
\emlineto{27.935}{18.453} 
\emlineto{27.935}{18.463} 
\emmoveto{27.935}{18.453} 
\emlineto{28.525}{21.351} 
\emlineto{28.525}{21.361} 
\emmoveto{28.525}{21.351} 
\emlineto{29.115}{19.203} 
\emlineto{29.115}{19.213} 
\emmoveto{29.115}{19.203} 
\emlineto{29.705}{20.091} 
\emlineto{29.705}{20.101} 
\emmoveto{29.705}{20.091} 
\emlineto{30.295}{20.146} 
\emlineto{30.295}{20.156} 
\emmoveto{30.295}{20.146} 
\emlineto{30.885}{18.156} 
\emlineto{30.885}{18.166} 
\emmoveto{30.885}{18.156} 
\emlineto{31.475}{23.448} 
\emlineto{31.475}{23.458} 
\emmoveto{31.475}{23.448} 
\emlineto{32.065}{24.232} 
\emlineto{32.065}{24.242} 
\emmoveto{32.065}{24.232} 
\emlineto{32.655}{20.144} 
\emlineto{32.655}{20.154} 
\emmoveto{32.655}{20.144} 
\emlineto{33.245}{22.941} 
\emlineto{33.245}{22.951} 
\emmoveto{33.245}{22.941} 
\emlineto{33.835}{17.914} 
\emlineto{33.835}{17.924} 
\emmoveto{33.835}{17.914} 
\emlineto{34.425}{22.770} 
\emlineto{34.425}{22.780} 
\emmoveto{34.425}{22.770} 
\emlineto{35.015}{17.266} 
\emlineto{35.015}{17.276} 
\emmoveto{35.015}{17.266} 
\emlineto{35.605}{16.375} 
\emlineto{35.605}{16.385} 
\emmoveto{35.605}{16.375} 
\emlineto{36.195}{18.930} 
\emlineto{36.195}{18.940} 
\emmoveto{36.195}{18.930} 
\emlineto{36.785}{23.816} 
\emlineto{36.785}{23.826} 
\emmoveto{36.785}{23.816} 
\emlineto{37.375}{22.541} 
\emmoveto{37.375}{22.531} 
\emlineto{37.965}{21.111} 
\emlineto{37.965}{21.121} 
\emmoveto{37.965}{21.111} 
\emlineto{38.555}{21.011} 
\emlineto{38.555}{21.021} 
\emmoveto{38.555}{21.011} 
\emlineto{39.145}{25.556} 
\emlineto{39.145}{25.566} 
\emmoveto{39.145}{25.556} 
\emlineto{39.735}{20.419} 
\emlineto{39.735}{20.429} 
\emmoveto{39.735}{20.419} 
\emlineto{40.324}{22.116} 
\emlineto{40.324}{22.126} 
\emmoveto{40.324}{22.116} 
\emlineto{40.914}{29.179} 
\emlineto{40.914}{29.189} 
\emmoveto{40.914}{29.179} 
\emlineto{41.504}{21.899} 
\emlineto{41.504}{21.909} 
\emmoveto{41.504}{21.899} 
\emlineto{42.094}{26.999} 
\emlineto{42.094}{27.009} 
\emmoveto{42.094}{26.999} 
\emlineto{42.684}{19.973} 
\emlineto{42.684}{19.983} 
\emmoveto{42.684}{19.973} 
\emlineto{43.274}{24.743} 
\emlineto{43.274}{24.753} 
\emmoveto{43.274}{24.743} 
\emlineto{43.864}{25.730} 
\emlineto{43.864}{25.740} 
\emmoveto{43.864}{25.730} 
\emlineto{44.454}{17.654} 
\emlineto{44.454}{17.664} 
\emmoveto{44.454}{17.654} 
\emlineto{45.044}{20.171} 
\emlineto{45.044}{20.181} 
\emmoveto{45.044}{20.171} 
\emlineto{45.634}{24.397} 
\emlineto{45.634}{24.407} 
\emmoveto{45.634}{24.397} 
\emlineto{46.224}{24.760} 
\emlineto{46.224}{24.770} 
\emmoveto{46.224}{24.760} 
\emlineto{46.814}{27.051} 
\emlineto{46.814}{27.061} 
\emmoveto{46.814}{27.051} 
\emlineto{47.404}{23.237} 
\emlineto{47.404}{23.247} 
\emmoveto{47.404}{23.237} 
\emlineto{47.994}{25.290} 
\emlineto{47.994}{25.300} 
\emmoveto{47.994}{25.290} 
\emlineto{48.584}{20.925} 
\emlineto{48.584}{20.935} 
\emmoveto{48.584}{20.925} 
\emlineto{49.174}{24.718} 
\emlineto{49.174}{24.728} 
\emmoveto{49.174}{24.718} 
\emlineto{49.764}{30.273} 
\emlineto{49.764}{30.283} 
\emmoveto{49.764}{30.273} 
\emlineto{50.354}{26.343} 
\emlineto{50.354}{26.353} 
\emmoveto{50.354}{26.343} 
\emlineto{50.944}{29.191} 
\emlineto{50.944}{29.201} 
\emmoveto{50.944}{29.191} 
\emlineto{51.534}{23.684} 
\emlineto{51.534}{23.694} 
\emmoveto{51.534}{23.684} 
\emlineto{52.124}{25.398} 
\emlineto{52.124}{25.408} 
\emmoveto{52.124}{25.398} 
\emlineto{52.714}{29.766} 
\emlineto{52.714}{29.776} 
\emmoveto{52.714}{29.766} 
\emlineto{53.304}{20.856} 
\emlineto{53.304}{20.866} 
\emmoveto{53.304}{20.856} 
\emlineto{53.894}{24.314} 
\emlineto{53.894}{24.324} 
\emmoveto{53.894}{24.314} 
\emlineto{54.484}{20.455} 
\emlineto{54.484}{20.465} 
\emmoveto{54.484}{20.455} 
\emlineto{55.074}{26.136} 
\emlineto{55.074}{26.146} 
\emmoveto{55.074}{26.136} 
\emlineto{55.664}{28.539} 
\emlineto{55.664}{28.549} 
\emmoveto{55.664}{28.539} 
\emlineto{56.254}{25.014} 
\emlineto{56.254}{25.024} 
\emmoveto{56.254}{25.014} 
\emlineto{56.844}{24.845} 
\emlineto{56.844}{24.855} 
\emmoveto{56.844}{24.845} 
\emlineto{57.434}{25.343} 
\emlineto{57.434}{25.353} 
\emmoveto{57.434}{25.343} 
\emlineto{58.024}{25.327} 
\emlineto{58.024}{25.337} 
\emmoveto{58.024}{25.327} 
\emlineto{58.614}{29.628} 
\emlineto{58.614}{29.638} 
\emmoveto{58.614}{29.628} 
\emlineto{59.204}{29.972} 
\emlineto{59.204}{29.982} 
\emmoveto{59.204}{29.972} 
\emlineto{59.794}{26.912} 
\emlineto{59.794}{26.922} 
\emmoveto{59.794}{26.912} 
\emlineto{60.383}{29.859} 
\emlineto{60.383}{29.869} 
\emmoveto{60.383}{29.859} 
\emlineto{60.973}{26.214} 
\emlineto{60.973}{26.224} 
\emmoveto{60.973}{26.214} 
\emlineto{61.563}{29.549} 
\emlineto{61.563}{29.559} 
\emmoveto{61.563}{29.549} 
\emlineto{62.153}{23.551} 
\emlineto{62.153}{23.561} 
\emmoveto{62.153}{23.551} 
\emlineto{62.743}{23.684} 
\emlineto{62.743}{23.694} 
\emmoveto{62.743}{23.684} 
\emlineto{63.333}{27.599} 
\emlineto{63.333}{27.609} 
\emmoveto{63.333}{27.599} 
\emlineto{63.923}{29.095} 
\emlineto{63.923}{29.105} 
\emmoveto{63.923}{29.095} 
\emlineto{64.513}{28.958} 
\emlineto{64.513}{28.968} 
\emmoveto{64.513}{28.958} 
\emlineto{65.103}{29.070} 
\emlineto{65.103}{29.080} 
\emmoveto{65.103}{29.070} 
\emlineto{65.693}{28.874} 
\emlineto{65.693}{28.884} 
\emmoveto{65.693}{28.874} 
\emlineto{66.283}{30.685} 
\emlineto{66.283}{30.695} 
\emmoveto{66.283}{30.685} 
\emlineto{66.873}{29.935} 
\emlineto{66.873}{29.945} 
\emmoveto{66.873}{29.935} 
\emlineto{67.463}{34.360} 
\emlineto{67.463}{34.370} 
\emmoveto{67.463}{34.360} 
\emlineto{68.053}{34.171} 
\emlineto{68.053}{34.181} 
\emmoveto{68.053}{34.171} 
\emlineto{68.643}{32.013} 
\emlineto{68.643}{32.023} 
\emmoveto{68.643}{32.013} 
\emlineto{69.233}{33.155} 
\emlineto{69.233}{33.165} 
\emmoveto{69.233}{33.155} 
\emlineto{69.823}{27.945} 
\emlineto{69.823}{27.955} 
\emmoveto{69.823}{27.945} 
\emlineto{70.413}{32.171} 
\emlineto{70.413}{32.181} 
\emmoveto{70.413}{32.171} 
\emlineto{71.003}{27.799} 
\emlineto{71.003}{27.809} 
\emmoveto{71.003}{27.799} 
\emlineto{71.593}{25.673} 
\emlineto{71.593}{25.683} 
\emmoveto{71.593}{25.673} 
\emlineto{72.183}{26.496} 
\emlineto{72.183}{26.506} 
\emmoveto{72.183}{26.496} 
\emlineto{72.773}{33.512} 
\emlineto{72.773}{33.522} 
\emmoveto{72.773}{33.512} 
\emlineto{73.363}{31.366} 
\emlineto{73.363}{31.376} 
\emmoveto{73.363}{31.366} 
\emlineto{73.953}{32.012} 
\emlineto{73.953}{32.022} 
\emmoveto{73.953}{32.012} 
\emlineto{74.543}{31.828} 
\emlineto{74.543}{31.838} 
\emmoveto{74.543}{31.828} 
\emlineto{75.133}{32.589} 
\emlineto{75.133}{32.599} 
\emmoveto{75.133}{32.589} 
\emlineto{75.723}{29.096} 
\emlineto{75.723}{29.106} 
\emmoveto{75.723}{29.096} 
\emlineto{76.313}{29.973} 
\emlineto{76.313}{29.983} 
\emmoveto{76.313}{29.973} 
\emlineto{76.903}{37.567} 
\emlineto{76.903}{37.577} 
\emmoveto{76.903}{37.567} 
\emlineto{77.493}{31.141} 
\emlineto{77.493}{31.151} 
\emmoveto{77.493}{31.141} 
\emlineto{78.083}{34.906} 
\emlineto{78.083}{34.916} 
\emmoveto{78.083}{34.906} 
\emlineto{78.673}{27.961} 
\emlineto{78.673}{27.971} 
\emmoveto{78.673}{27.961} 
\emlineto{79.263}{33.105} 
\emlineto{79.263}{33.115} 
\emmoveto{79.263}{33.105} 
\emlineto{79.853}{36.483} 
\emlineto{79.853}{36.493} 
\emmoveto{79.853}{36.483} 
\emlineto{80.442}{26.586} 
\emlineto{80.442}{26.596} 
\emmoveto{80.442}{26.586} 
\emlineto{81.032}{29.001} 
\emlineto{81.032}{29.011} 
\emmoveto{81.032}{29.001} 
\emlineto{81.622}{31.833} 
\emmoveto{81.622}{31.823} 
\emlineto{82.212}{34.359} 
\emlineto{82.212}{34.369} 
\emmoveto{82.212}{34.359} 
\emlineto{82.802}{35.242} 
\emlineto{82.802}{35.252} 
\emmoveto{82.802}{35.242} 
\emlineto{83.392}{32.812} 
\emlineto{83.392}{32.822} 
\emmoveto{83.392}{32.812} 
\emlineto{83.982}{34.435} 
\emlineto{83.982}{34.445} 
\emmoveto{83.982}{34.435} 
\emlineto{84.572}{31.658} 
\emlineto{84.572}{31.668} 
\emmoveto{84.572}{31.658} 
\emlineto{85.162}{32.703} 
\emlineto{85.162}{32.713} 
\emmoveto{85.162}{32.703} 
\emlineto{85.752}{37.679} 
\emlineto{85.752}{37.689} 
\emmoveto{85.752}{37.679} 
\emlineto{86.342}{36.813} 
\emlineto{86.342}{36.823} 
\emmoveto{86.342}{36.813} 
\emlineto{86.932}{36.189} 
\emlineto{86.932}{36.199} 
\emmoveto{86.932}{36.189} 
\emlineto{87.522}{34.932} 
\emlineto{87.522}{34.942} 
\emmoveto{87.522}{34.932} 
\emlineto{88.112}{34.586} 
\emlineto{88.112}{34.596} 
\emmoveto{88.112}{34.586} 
\emlineto{88.702}{37.560} 
\emlineto{88.702}{37.570} 
\emmoveto{88.702}{37.560} 
\emlineto{89.292}{30.727} 
\emlineto{89.292}{30.737} 
\emmoveto{89.292}{30.727} 
\emlineto{89.882}{33.759} 
\emlineto{89.882}{33.769} 
\emmoveto{89.882}{33.759} 
\emlineto{90.472}{32.709} 
\emlineto{90.472}{32.719} 
\emmoveto{90.472}{32.709} 
\emlineto{91.062}{34.458} 
\emlineto{91.062}{34.468} 
\emmoveto{91.062}{34.458} 
\emlineto{91.652}{37.703} 
\emlineto{91.652}{37.713} 
\emmoveto{91.652}{37.703} 
\emlineto{92.242}{35.805} 
\emlineto{92.242}{35.815} 
\emmoveto{92.242}{35.805} 
\emlineto{92.832}{37.095} 
\emlineto{92.832}{37.105} 
\emmoveto{92.832}{37.095} 
\emlineto{93.422}{36.708} 
\emlineto{93.422}{36.718} 
\emmoveto{93.422}{36.708} 
\emlineto{94.012}{34.683} 
\emlineto{94.012}{34.693} 
\emmoveto{94.012}{34.683} 
\emlineto{94.602}{42.172} 
\emlineto{94.602}{42.182} 
\emmoveto{94.602}{42.172} 
\emlineto{95.192}{41.102} 
\emlineto{95.192}{41.112} 
\emmoveto{95.192}{41.102} 
\emlineto{95.782}{36.159} 
\emlineto{95.782}{36.169} 
\emmoveto{95.782}{36.159} 
\emlineto{96.372}{38.144} 
\emlineto{96.372}{38.154} 
\emmoveto{96.372}{38.144} 
\emlineto{96.962}{34.480} 
\emlineto{96.962}{34.490} 
\emmoveto{96.962}{34.480} 
\emlineto{97.552}{39.326} 
\emlineto{97.552}{39.336} 
\emmoveto{97.552}{39.326} 
\emlineto{98.142}{34.630} 
\emlineto{98.142}{34.640} 
\emmoveto{98.142}{34.630} 
\emlineto{98.732}{33.908} 
\emlineto{98.732}{33.918} 
\emmoveto{98.732}{33.908} 
\emlineto{99.322}{34.625} 
\emlineto{99.322}{34.635} 
\emmoveto{99.322}{34.625} 
\emlineto{99.912}{40.138} 
\emlineto{99.912}{40.148} 
\emmoveto{99.912}{40.138} 
\emlineto{100.501}{39.688} 
\emlineto{100.501}{39.698} 
\emmoveto{100.501}{39.688} 
\emlineto{101.091}{37.965} 
\emlineto{101.091}{37.975} 
\emmoveto{101.091}{37.965} 
\emlineto{101.681}{36.524} 
\emlineto{101.681}{36.534} 
\emmoveto{101.681}{36.524} 
\emlineto{102.271}{41.282} 
\emlineto{102.271}{41.292} 
\emmoveto{102.271}{41.282} 
\emlineto{102.861}{38.337} 
\emlineto{102.861}{38.347} 
\emmoveto{102.861}{38.337} 
\emlineto{103.451}{39.856} 
\emlineto{103.451}{39.866} 
\emmoveto{103.451}{39.856} 
\emlineto{104.041}{45.577} 
\emlineto{104.041}{45.587} 
\emmoveto{104.041}{45.577} 
\emlineto{104.631}{39.034} 
\emlineto{104.631}{39.044} 
\emmoveto{104.631}{39.034} 
\emlineto{105.221}{43.202} 
\emlineto{105.221}{43.212} 
\emmoveto{105.221}{43.202} 
\emlineto{105.811}{36.463} 
\emlineto{105.811}{36.473} 
\emmoveto{105.811}{36.463} 
\emlineto{106.401}{40.251} 
\emlineto{106.401}{40.261} 
\emmoveto{106.401}{40.251} 
\emlineto{106.991}{41.980} 
\emlineto{106.991}{41.990} 
\emmoveto{106.991}{41.980} 
\emlineto{107.581}{35.548} 
\emlineto{107.581}{35.558} 
\emmoveto{107.581}{35.548} 
\emlineto{108.171}{34.709} 
\emlineto{108.171}{34.719} 
\emmoveto{108.171}{34.709} 
\emlineto{108.761}{42.257} 
\emlineto{108.761}{42.267} 
\emmoveto{108.761}{42.257} 
\emlineto{109.351}{41.478} 
\emlineto{109.351}{41.488} 
\emmoveto{109.351}{41.478} 
\emlineto{109.941}{43.974} 
\emlineto{109.941}{43.984} 
\emmoveto{109.941}{43.974} 
\emlineto{110.531}{40.245} 
\emlineto{110.531}{40.255} 
\emmoveto{110.531}{40.245} 
\emlineto{111.121}{43.105} 
\emlineto{111.121}{43.115} 
\emmoveto{111.121}{43.105} 
\emlineto{111.711}{39.427} 
\emlineto{111.711}{39.437} 
\emmoveto{111.711}{39.427} 
\emlineto{112.301}{41.249} 
\emlineto{112.301}{41.259} 
\emmoveto{112.301}{41.249} 
\emlineto{112.891}{47.503} 
\emlineto{112.891}{47.513} 
\emmoveto{112.891}{47.503} 
\emlineto{113.481}{43.225} 
\emlineto{113.481}{43.235} 
\emmoveto{113.481}{43.225} 
\emlineto{114.071}{46.078} 
\emlineto{114.071}{46.088} 
\emmoveto{114.071}{46.078} 
\emlineto{114.661}{40.503} 
\emlineto{114.661}{40.513} 
\emmoveto{114.661}{40.503} 
\emlineto{115.251}{42.428} 
\emlineto{115.251}{42.438} 
\emmoveto{115.251}{42.428} 
\emlineto{115.841}{45.986} 
\emlineto{115.841}{45.996} 
\emmoveto{115.841}{45.986} 
\emlineto{116.431}{37.252} 
\emlineto{116.431}{37.262} 
\emmoveto{116.431}{37.252} 
\emlineto{117.021}{41.367} 
\emlineto{117.021}{41.377} 
\emmoveto{117.021}{41.367} 
\emlineto{117.611}{38.603} 
\emlineto{117.611}{38.613} 
\emmoveto{117.611}{38.603} 
\emlineto{118.201}{42.240} 
\emlineto{118.201}{42.250} 
\emmoveto{118.201}{42.240} 
\emlineto{118.791}{45.003} 
\emlineto{118.791}{45.013} 
\emmoveto{118.791}{45.003} 
\emlineto{119.381}{41.894} 
\emlineto{119.381}{41.904} 
\emmoveto{119.381}{41.894} 
\emlineto{119.971}{40.151} 
\emlineto{119.971}{40.161} 
\emmoveto{119.971}{40.151} 
\emlineto{120.560}{42.420} 
\emlineto{120.560}{42.430} 
\emmoveto{120.560}{42.420} 
\emlineto{121.150}{40.635} 
\emlineto{121.150}{40.645} 
\emmoveto{121.150}{40.635} 
\emlineto{121.740}{47.690} 
\emlineto{121.740}{47.700} 
\emmoveto{121.740}{47.690} 
\emlineto{122.330}{46.775} 
\emlineto{122.330}{46.785} 
\emmoveto{122.330}{46.775} 
\emlineto{122.920}{44.187} 
\emlineto{122.920}{44.197} 
\emmoveto{122.920}{44.187} 
\emlineto{123.510}{46.539} 
\emlineto{123.510}{46.549} 
\emmoveto{123.510}{46.539} 
\emlineto{124.100}{43.577} 
\emlineto{124.100}{43.587} 
\emmoveto{124.100}{43.577} 
\emlineto{124.690}{46.634} 
\emlineto{124.690}{46.644} 
\emmoveto{124.690}{46.634} 
\emlineto{125.280}{41.129} 
\emlineto{125.280}{41.139} 
\emmoveto{125.280}{41.129} 
\emlineto{125.870}{41.360} 
\emlineto{125.870}{41.370} 
\emmoveto{125.870}{41.360} 
\emlineto{126.460}{45.177} 
\emlineto{126.460}{45.187} 
\emmoveto{126.460}{45.177} 
\emlineto{127.050}{45.461} 
\emlineto{127.050}{45.471} 
\emmoveto{127.050}{45.461} 
\emlineto{127.640}{45.181} 
\emlineto{127.640}{45.191} 
\emmoveto{127.640}{45.181} 
\emlineto{128.230}{45.308} 
\emlineto{128.230}{45.318} 
\emmoveto{128.230}{45.308} 
\emlineto{128.820}{45.802} 
\emlineto{128.820}{45.812} 
\emmoveto{128.820}{45.802} 
\emlineto{129.410}{47.729} 
\emlineto{129.410}{47.739} 
\emshow{24.980}{25.400}{} 
\emmoveto{12.000}{14.073} 
\emlineto{12.596}{14.409} 
\emmoveto{12.596}{14.399} 
\emlineto{13.186}{14.730} 
\emmoveto{13.186}{14.720} 
\emlineto{13.776}{15.052} 
\emmoveto{13.776}{15.042} 
\emlineto{14.366}{15.374} 
\emmoveto{14.366}{15.364} 
\emlineto{14.956}{15.696} 
\emmoveto{14.956}{15.686} 
\emlineto{15.546}{16.017} 
\emmoveto{15.546}{16.007} 
\emlineto{16.136}{16.339} 
\emmoveto{16.136}{16.329} 
\emlineto{16.726}{16.661} 
\emmoveto{16.726}{16.651} 
\emlineto{17.316}{16.982} 
\emmoveto{17.316}{16.972} 
\emlineto{17.906}{17.304} 
\emmoveto{17.906}{17.294} 
\emlineto{18.496}{17.627} 
\emmoveto{18.496}{17.617} 
\emlineto{19.086}{17.949} 
\emmoveto{19.086}{17.939} 
\emlineto{19.676}{18.271} 
\emmoveto{19.676}{18.261} 
\emlineto{20.265}{18.592} 
\emmoveto{20.265}{18.582} 
\emlineto{20.855}{18.914} 
\emmoveto{20.855}{18.904} 
\emlineto{21.445}{19.236} 
\emmoveto{21.445}{19.226} 
\emlineto{22.035}{19.558} 
\emmoveto{22.035}{19.548} 
\emlineto{22.625}{19.879} 
\emmoveto{22.625}{19.869} 
\emlineto{23.215}{20.201} 
\emmoveto{23.215}{20.191} 
\emlineto{23.805}{20.523} 
\emmoveto{23.805}{20.513} 
\emlineto{24.395}{20.846} 
\emmoveto{24.395}{20.836} 
\emlineto{24.985}{21.167} 
\emmoveto{24.985}{21.157} 
\emlineto{25.575}{21.489} 
\emmoveto{25.575}{21.479} 
\emlineto{26.165}{21.811} 
\emmoveto{26.165}{21.801} 
\emlineto{26.755}{22.133} 
\emmoveto{26.755}{22.123} 
\emlineto{27.345}{22.454} 
\emmoveto{27.345}{22.444} 
\emlineto{27.935}{22.776} 
\emmoveto{27.935}{22.766} 
\emlineto{28.525}{23.098} 
\emmoveto{28.525}{23.088} 
\emlineto{29.115}{23.420} 
\emmoveto{29.115}{23.410} 
\emlineto{29.705}{23.741} 
\emmoveto{29.705}{23.731} 
\emlineto{30.295}{24.064} 
\emmoveto{30.295}{24.054} 
\emlineto{30.885}{24.386} 
\emmoveto{30.885}{24.376} 
\emlineto{31.475}{24.708} 
\emmoveto{31.475}{24.698} 
\emlineto{32.065}{25.029} 
\emmoveto{32.065}{25.019} 
\emlineto{32.655}{25.351} 
\emmoveto{32.655}{25.341} 
\emlineto{33.245}{25.673} 
\emmoveto{33.245}{25.663} 
\emlineto{33.835}{25.995} 
\emmoveto{33.835}{25.985} 
\emlineto{34.425}{26.316} 
\emmoveto{34.425}{26.306} 
\emlineto{35.015}{26.638} 
\emmoveto{35.015}{26.628} 
\emlineto{35.605}{26.960} 
\emmoveto{35.605}{26.950} 
\emlineto{36.195}{27.283} 
\emmoveto{36.195}{27.273} 
\emlineto{36.785}{27.605} 
\emmoveto{36.785}{27.595} 
\emlineto{37.375}{27.926} 
\emmoveto{37.375}{27.916} 
\emlineto{37.965}{28.248} 
\emmoveto{37.965}{28.238} 
\emlineto{38.555}{28.570} 
\emmoveto{38.555}{28.560} 
\emlineto{39.145}{28.892} 
\emmoveto{39.145}{28.882} 
\emlineto{39.735}{29.213} 
\emmoveto{39.735}{29.203} 
\emlineto{40.324}{29.535} 
\emmoveto{40.324}{29.525} 
\emlineto{40.914}{29.857} 
\emmoveto{40.914}{29.847} 
\emlineto{41.504}{30.178} 
\emmoveto{41.504}{30.168} 
\emlineto{42.094}{30.501} 
\emmoveto{42.094}{30.491} 
\emlineto{42.684}{30.823} 
\emmoveto{42.684}{30.813} 
\emlineto{43.274}{31.145} 
\emmoveto{43.274}{31.135} 
\emlineto{43.864}{31.467} 
\emmoveto{43.864}{31.457} 
\emlineto{44.454}{31.788} 
\emmoveto{44.454}{31.778} 
\emlineto{45.044}{32.110} 
\emmoveto{45.044}{32.100} 
\emlineto{45.634}{32.432} 
\emmoveto{45.634}{32.422} 
\emlineto{46.224}{32.754} 
\emmoveto{46.224}{32.744} 
\emlineto{46.814}{33.075} 
\emmoveto{46.814}{33.065} 
\emlineto{47.404}{33.397} 
\emmoveto{47.404}{33.387} 
\emlineto{47.994}{33.720} 
\emmoveto{47.994}{33.710} 
\emlineto{48.584}{34.042} 
\emmoveto{48.584}{34.032} 
\emlineto{49.174}{34.363} 
\emmoveto{49.174}{34.353} 
\emlineto{49.764}{34.685} 
\emmoveto{49.764}{34.675} 
\emlineto{50.354}{35.007} 
\emmoveto{50.354}{34.997} 
\emlineto{50.944}{35.329} 
\emmoveto{50.944}{35.319} 
\emlineto{51.534}{35.650} 
\emmoveto{51.534}{35.640} 
\emlineto{52.124}{35.972} 
\emmoveto{52.124}{35.962} 
\emlineto{52.714}{36.294} 
\emmoveto{52.714}{36.284} 
\emlineto{53.304}{36.616} 
\emmoveto{53.304}{36.606} 
\emlineto{53.894}{36.939} 
\emmoveto{53.894}{36.929} 
\emlineto{54.484}{37.260} 
\emmoveto{54.484}{37.250} 
\emlineto{55.074}{37.582} 
\emmoveto{55.074}{37.572} 
\emlineto{55.664}{37.904} 
\emmoveto{55.664}{37.894} 
\emlineto{56.254}{38.225} 
\emmoveto{56.254}{38.215} 
\emlineto{56.844}{38.547} 
\emmoveto{56.844}{38.537} 
\emlineto{57.434}{38.869} 
\emmoveto{57.434}{38.859} 
\emlineto{58.024}{39.191} 
\emmoveto{58.024}{39.181} 
\emlineto{58.614}{39.512} 
\emmoveto{58.614}{39.502} 
\emlineto{59.204}{39.834} 
\emmoveto{59.204}{39.824} 
\emlineto{59.794}{40.157} 
\emmoveto{59.794}{40.147} 
\emlineto{60.383}{40.479} 
\emmoveto{60.383}{40.469} 
\emlineto{60.973}{40.801} 
\emmoveto{60.973}{40.791} 
\emlineto{61.563}{41.122} 
\emmoveto{61.563}{41.112} 
\emlineto{62.153}{41.444} 
\emmoveto{62.153}{41.434} 
\emlineto{62.743}{41.766} 
\emmoveto{62.743}{41.756} 
\emlineto{63.333}{42.087} 
\emmoveto{63.333}{42.077} 
\emlineto{63.923}{42.409} 
\emmoveto{63.923}{42.399} 
\emlineto{64.513}{42.731} 
\emmoveto{64.513}{42.721} 
\emlineto{65.103}{43.053} 
\emmoveto{65.103}{43.043} 
\emlineto{65.693}{43.376} 
\emmoveto{65.693}{43.366} 
\emlineto{66.283}{43.697} 
\emmoveto{66.283}{43.687} 
\emlineto{66.873}{44.019} 
\emmoveto{66.873}{44.009} 
\emlineto{67.463}{44.341} 
\emmoveto{67.463}{44.331} 
\emlineto{68.053}{44.663} 
\emmoveto{68.053}{44.653} 
\emlineto{68.643}{44.984} 
\emmoveto{68.643}{44.974} 
\emlineto{69.233}{45.306} 
\emmoveto{69.233}{45.296} 
\emlineto{69.823}{45.628} 
\emmoveto{69.823}{45.618} 
\emlineto{70.413}{45.949} 
\emmoveto{70.413}{45.939} 
\emlineto{71.003}{46.271} 
\emmoveto{71.003}{46.261} 
\emlineto{71.593}{46.594} 
\emmoveto{71.593}{46.584} 
\emlineto{72.183}{46.916} 
\emmoveto{72.183}{46.906} 
\emlineto{72.773}{47.238} 
\emmoveto{72.773}{47.228} 
\emlineto{73.363}{47.559} 
\emmoveto{73.363}{47.549} 
\emlineto{73.953}{47.881} 
\emmoveto{73.953}{47.871} 
\emlineto{74.543}{48.203} 
\emmoveto{74.543}{48.193} 
\emlineto{75.133}{48.525} 
\emmoveto{75.133}{48.515} 
\emlineto{75.723}{48.846} 
\emmoveto{75.723}{48.836} 
\emlineto{76.313}{49.168} 
\emmoveto{76.313}{49.158} 
\emlineto{76.903}{49.490} 
\emmoveto{76.903}{49.480} 
\emlineto{77.493}{49.813} 
\emmoveto{77.493}{49.803} 
\emlineto{78.083}{50.135} 
\emmoveto{78.083}{50.125} 
\emlineto{78.673}{50.456} 
\emmoveto{78.673}{50.446} 
\emlineto{79.263}{50.778} 
\emmoveto{79.263}{50.768} 
\emlineto{79.853}{51.100} 
\emmoveto{79.853}{51.090} 
\emlineto{80.442}{51.421} 
\emmoveto{80.442}{51.411} 
\emlineto{81.032}{51.743} 
\emmoveto{81.032}{51.733} 
\emlineto{81.622}{52.065} 
\emmoveto{81.622}{52.055} 
\emlineto{82.212}{52.387} 
\emmoveto{82.212}{52.377} 
\emlineto{82.802}{52.708} 
\emmoveto{82.802}{52.698} 
\emlineto{83.392}{53.031} 
\emmoveto{83.392}{53.021} 
\emlineto{83.982}{53.353} 
\emmoveto{83.982}{53.343} 
\emlineto{84.572}{53.675} 
\emmoveto{84.572}{53.665} 
\emlineto{85.162}{53.997} 
\emmoveto{85.162}{53.987} 
\emlineto{85.752}{54.318} 
\emmoveto{85.752}{54.308} 
\emlineto{86.342}{54.640} 
\emmoveto{86.342}{54.630} 
\emlineto{86.932}{54.962} 
\emmoveto{86.932}{54.952} 
\emlineto{87.522}{55.283} 
\emmoveto{87.522}{55.273} 
\emlineto{88.112}{55.605} 
\emmoveto{88.112}{55.595} 
\emlineto{88.702}{55.927} 
\emmoveto{88.702}{55.917} 
\emlineto{89.292}{56.250} 
\emmoveto{89.292}{56.240} 
\emlineto{89.882}{56.572} 
\emmoveto{89.882}{56.562} 
\emlineto{90.472}{56.893} 
\emmoveto{90.472}{56.883} 
\emlineto{91.062}{57.215} 
\emmoveto{91.062}{57.205} 
\emlineto{91.652}{57.537} 
\emmoveto{91.652}{57.527} 
\emlineto{92.242}{57.859} 
\emmoveto{92.242}{57.849} 
\emlineto{92.832}{58.180} 
\emmoveto{92.832}{58.170} 
\emlineto{93.422}{58.502} 
\emmoveto{93.422}{58.492} 
\emlineto{94.012}{58.824} 
\emmoveto{94.012}{58.814} 
\emlineto{94.602}{59.145} 
\emmoveto{94.602}{59.135} 
\emlineto{95.192}{59.468} 
\emmoveto{95.192}{59.458} 
\emlineto{95.782}{59.790} 
\emmoveto{95.782}{59.780} 
\emlineto{96.372}{60.112} 
\emmoveto{96.372}{60.102} 
\emlineto{96.962}{60.434} 
\emmoveto{96.962}{60.424} 
\emlineto{97.552}{60.755} 
\emmoveto{97.552}{60.745} 
\emlineto{98.142}{61.077} 
\emmoveto{98.142}{61.067} 
\emlineto{98.732}{61.399} 
\emmoveto{98.732}{61.389} 
\emlineto{99.322}{61.721} 
\emmoveto{99.322}{61.711} 
\emlineto{99.912}{62.042} 
\emmoveto{99.912}{62.032} 
\emlineto{100.501}{62.364} 
\emmoveto{100.501}{62.354} 
\emlineto{101.091}{62.687} 
\emmoveto{101.091}{62.677} 
\emlineto{101.681}{63.009} 
\emmoveto{101.681}{62.999} 
\emlineto{102.271}{63.330} 
\emmoveto{102.271}{63.320} 
\emlineto{102.861}{63.652} 
\emmoveto{102.861}{63.642} 
\emlineto{103.451}{63.974} 
\emmoveto{103.451}{63.964} 
\emlineto{104.041}{64.296} 
\emmoveto{104.041}{64.286} 
\emlineto{104.631}{64.617} 
\emmoveto{104.631}{64.607} 
\emlineto{105.221}{64.939} 
\emmoveto{105.221}{64.929} 
\emlineto{105.811}{65.261} 
\emmoveto{105.811}{65.251} 
\emlineto{106.401}{65.583} 
\emmoveto{106.401}{65.573} 
\emlineto{106.991}{65.906} 
\emmoveto{106.991}{65.896} 
\emlineto{107.581}{66.227} 
\emmoveto{107.581}{66.217} 
\emlineto{108.171}{66.549} 
\emmoveto{108.171}{66.539} 
\emlineto{108.761}{66.871} 
\emmoveto{108.761}{66.861} 
\emlineto{109.351}{67.192} 
\emmoveto{109.351}{67.182} 
\emlineto{109.941}{67.514} 
\emmoveto{109.941}{67.504} 
\emlineto{110.531}{67.836} 
\emmoveto{110.531}{67.826} 
\emlineto{111.121}{68.158} 
\emmoveto{111.121}{68.148} 
\emlineto{111.711}{68.479} 
\emmoveto{111.711}{68.469} 
\emlineto{112.301}{68.801} 
\emmoveto{112.301}{68.791} 
\emlineto{112.891}{69.124} 
\emmoveto{112.891}{69.114} 
\emlineto{113.481}{69.446} 
\emmoveto{113.481}{69.436} 
\emlineto{114.071}{69.768} 
\emmoveto{114.071}{69.758} 
\emlineto{114.661}{70.089} 
\emmoveto{114.661}{70.079} 
\emlineto{115.251}{70.411} 
\emmoveto{115.251}{70.401} 
\emlineto{115.841}{70.733} 
\emmoveto{115.841}{70.723} 
\emlineto{116.431}{71.054} 
\emmoveto{116.431}{71.044} 
\emlineto{117.021}{71.376} 
\emmoveto{117.021}{71.366} 
\emlineto{117.611}{71.698} 
\emmoveto{117.611}{71.688} 
\emlineto{118.201}{72.020} 
\emmoveto{118.201}{72.010} 
\emlineto{118.791}{72.343} 
\emmoveto{118.791}{72.333} 
\emlineto{119.381}{72.664} 
\emmoveto{119.381}{72.654} 
\emlineto{119.971}{72.986} 
\emmoveto{119.971}{72.976} 
\emlineto{120.560}{73.308} 
\emmoveto{120.560}{73.298} 
\emlineto{121.150}{73.630} 
\emmoveto{121.150}{73.620} 
\emlineto{121.740}{73.951} 
\emmoveto{121.740}{73.941} 
\emlineto{122.330}{74.273} 
\emmoveto{122.330}{74.263} 
\emlineto{122.920}{74.595} 
\emmoveto{122.920}{74.585} 
\emlineto{123.510}{74.916} 
\emmoveto{123.510}{74.906} 
\emlineto{124.100}{75.238} 
\emmoveto{124.100}{75.228} 
\emlineto{124.690}{75.561} 
\emmoveto{124.690}{75.551} 
\emlineto{125.280}{75.883} 
\emmoveto{125.280}{75.873} 
\emlineto{125.870}{76.205} 
\emmoveto{125.870}{76.195} 
\emlineto{126.460}{76.526} 
\emmoveto{126.460}{76.516} 
\emlineto{127.050}{76.848} 
\emmoveto{127.050}{76.838} 
\emlineto{127.640}{77.170} 
\emmoveto{127.640}{77.160} 
\emlineto{128.230}{77.492} 
\emmoveto{128.230}{77.482} 
\emlineto{128.820}{77.813} 
\emmoveto{128.820}{77.803} 
\emlineto{129.410}{78.135} 
\emshow{24.980}{25.400}{} 
\emmoveto{12.000}{14.073} 
\emlineto{12.596}{11.469} 
\emlineto{12.596}{11.479} 
\emmoveto{12.596}{11.469} 
\emlineto{13.186}{14.941} 
\emlineto{13.186}{14.951} 
\emmoveto{13.186}{14.941} 
\emlineto{13.776}{22.120} 
\emlineto{13.776}{22.130} 
\emmoveto{13.776}{22.120} 
\emlineto{14.366}{16.765} 
\emlineto{14.366}{16.775} 
\emmoveto{14.366}{16.765} 
\emlineto{14.956}{20.747} 
\emlineto{14.956}{20.757} 
\emmoveto{14.956}{20.747} 
\emlineto{15.546}{12.782} 
\emlineto{15.546}{12.792} 
\emmoveto{15.546}{12.782} 
\emlineto{16.136}{18.655} 
\emlineto{16.136}{18.665} 
\emmoveto{16.136}{18.655} 
\emlineto{16.726}{19.988} 
\emlineto{16.726}{19.998} 
\emmoveto{16.726}{19.988} 
\emlineto{17.316}{11.834} 
\emlineto{17.316}{11.844} 
\emmoveto{17.316}{11.834} 
\emlineto{17.906}{16.337} 
\emlineto{17.906}{16.347} 
\emmoveto{17.906}{16.337} 
\emlineto{18.496}{17.295} 
\emlineto{18.496}{17.305} 
\emmoveto{18.496}{17.295} 
\emlineto{19.086}{20.380} 
\emlineto{19.086}{20.390} 
\emmoveto{19.086}{20.380} 
\emlineto{19.676}{22.596} 
\emlineto{19.676}{22.606} 
\emmoveto{19.676}{22.596} 
\emlineto{20.265}{18.530} 
\emlineto{20.265}{18.540} 
\emmoveto{20.265}{18.530} 
\emlineto{20.855}{19.897} 
\emlineto{20.855}{19.907} 
\emmoveto{20.855}{19.897} 
\emlineto{21.445}{19.223} 
\emlineto{21.445}{19.233} 
\emmoveto{21.445}{19.223} 
\emlineto{22.035}{20.699} 
\emlineto{22.035}{20.709} 
\emmoveto{22.035}{20.699} 
\emlineto{22.625}{25.536} 
\emlineto{22.625}{25.546} 
\emmoveto{22.625}{25.536} 
\emlineto{23.215}{23.895} 
\emlineto{23.215}{23.905} 
\emmoveto{23.215}{23.895} 
\emlineto{23.805}{23.236} 
\emlineto{23.805}{23.246} 
\emmoveto{23.805}{23.236} 
\emlineto{24.395}{22.382} 
\emlineto{24.395}{22.392} 
\emmoveto{24.395}{22.382} 
\emlineto{24.985}{20.796} 
\emlineto{24.985}{20.806} 
\emmoveto{24.985}{20.796} 
\emlineto{25.575}{24.153} 
\emlineto{25.575}{24.163} 
\emmoveto{25.575}{24.153} 
\emlineto{26.165}{18.366} 
\emlineto{26.165}{18.376} 
\emmoveto{26.165}{18.366} 
\emlineto{26.755}{21.167} 
\emlineto{26.755}{21.177} 
\emmoveto{26.755}{21.167} 
\emlineto{27.345}{21.773} 
\emlineto{27.345}{21.783} 
\emmoveto{27.345}{21.773} 
\emlineto{27.935}{23.361} 
\emlineto{27.935}{23.371} 
\emmoveto{27.935}{23.361} 
\emlineto{28.525}{26.476} 
\emlineto{28.525}{26.486} 
\emmoveto{28.525}{26.476} 
\emlineto{29.115}{24.640} 
\emlineto{29.115}{24.650} 
\emmoveto{29.115}{24.640} 
\emlineto{29.705}{25.757} 
\emlineto{29.705}{25.767} 
\emmoveto{29.705}{25.757} 
\emlineto{30.295}{26.059} 
\emlineto{30.295}{26.069} 
\emmoveto{30.295}{26.059} 
\emlineto{30.885}{24.173} 
\emlineto{30.885}{24.183} 
\emmoveto{30.885}{24.173} 
\emlineto{31.475}{29.682} 
\emlineto{31.475}{29.692} 
\emmoveto{31.475}{29.682} 
\emlineto{32.065}{30.365} 
\emlineto{32.065}{30.375} 
\emmoveto{32.065}{30.365} 
\emlineto{32.655}{26.134} 
\emlineto{32.655}{26.144} 
\emmoveto{32.655}{26.134} 
\emlineto{33.245}{28.842} 
\emlineto{33.245}{28.852} 
\emmoveto{33.245}{28.842} 
\emlineto{33.835}{24.027} 
\emlineto{33.835}{24.037} 
\emmoveto{33.835}{24.027} 
\emlineto{34.425}{29.078} 
\emlineto{34.425}{29.088} 
\emmoveto{34.425}{29.078} 
\emlineto{35.015}{23.713} 
\emlineto{35.015}{23.723} 
\emmoveto{35.015}{23.713} 
\emlineto{35.605}{23.179} 
\emlineto{35.605}{23.189} 
\emmoveto{35.605}{23.179} 
\emlineto{36.195}{26.090} 
\emlineto{36.195}{26.100} 
\emmoveto{36.195}{26.090} 
\emlineto{36.785}{31.102} 
\emlineto{36.785}{31.112} 
\emmoveto{36.785}{31.102} 
\emlineto{37.375}{29.983} 
\emmoveto{37.375}{29.973} 
\emlineto{37.965}{28.868} 
\emlineto{37.965}{28.878} 
\emmoveto{37.965}{28.868} 
\emlineto{38.555}{29.012} 
\emlineto{38.555}{29.022} 
\emmoveto{38.555}{29.012} 
\emlineto{39.145}{33.826} 
\emlineto{39.145}{33.836} 
\emmoveto{39.145}{33.826} 
\emlineto{39.735}{28.805} 
\emlineto{39.735}{28.815} 
\emmoveto{39.735}{28.805} 
\emlineto{40.324}{30.710} 
\emlineto{40.324}{30.720} 
\emmoveto{40.324}{30.710} 
\emlineto{40.914}{37.791} 
\emlineto{40.914}{37.801} 
\emmoveto{40.914}{37.791} 
\emlineto{41.504}{30.295} 
\emlineto{41.504}{30.305} 
\emmoveto{41.504}{30.295} 
\emlineto{42.094}{35.312} 
\emlineto{42.094}{35.322} 
\emmoveto{42.094}{35.312} 
\emlineto{42.684}{28.402} 
\emlineto{42.684}{28.412} 
\emmoveto{42.684}{28.402} 
\emlineto{43.274}{33.408} 
\emlineto{43.274}{33.418} 
\emmoveto{43.274}{33.408} 
\emlineto{43.864}{34.516} 
\emlineto{43.864}{34.526} 
\emmoveto{43.864}{34.516} 
\emlineto{44.454}{26.745} 
\emlineto{44.454}{26.755} 
\emmoveto{44.454}{26.745} 
\emlineto{45.044}{29.605} 
\emlineto{45.044}{29.615} 
\emmoveto{45.044}{29.605} 
\emlineto{45.634}{34.077} 
\emlineto{45.634}{34.087} 
\emmoveto{45.634}{34.077} 
\emlineto{46.224}{34.516} 
\emlineto{46.224}{34.526} 
\emmoveto{46.224}{34.516} 
\emlineto{46.814}{37.116} 
\emlineto{46.814}{37.126} 
\emmoveto{46.814}{37.116} 
\emlineto{47.404}{33.578} 
\emlineto{47.404}{33.588} 
\emmoveto{47.404}{33.578} 
\emlineto{47.994}{35.908} 
\emlineto{47.994}{35.918} 
\emmoveto{47.994}{35.908} 
\emlineto{48.584}{31.665} 
\emlineto{48.584}{31.675} 
\emmoveto{48.584}{31.665} 
\emlineto{49.174}{35.627} 
\emlineto{49.174}{35.637} 
\emmoveto{49.174}{35.627} 
\emlineto{49.764}{41.346} 
\emlineto{49.764}{41.356} 
\emmoveto{49.764}{41.346} 
\emlineto{50.354}{37.165} 
\emlineto{50.354}{37.175} 
\emmoveto{50.354}{37.165} 
\emlineto{50.944}{39.965} 
\emlineto{50.944}{39.975} 
\emmoveto{50.944}{39.965} 
\emlineto{51.534}{34.382} 
\emlineto{51.534}{34.392} 
\emmoveto{51.534}{34.382} 
\emlineto{52.124}{36.414} 
\emlineto{52.124}{36.424} 
\emmoveto{52.124}{36.414} 
\emlineto{52.714}{40.934} 
\emlineto{52.714}{40.944} 
\emmoveto{52.714}{40.934} 
\emlineto{53.304}{32.223} 
\emlineto{53.304}{32.233} 
\emmoveto{53.304}{32.223} 
\emlineto{53.894}{36.029} 
\emlineto{53.894}{36.039} 
\emmoveto{53.894}{36.029} 
\emlineto{54.484}{32.501} 
\emlineto{54.484}{32.511} 
\emmoveto{54.484}{32.501} 
\emlineto{55.074}{38.259} 
\emlineto{55.074}{38.269} 
\emmoveto{55.074}{38.259} 
\emlineto{55.664}{40.904} 
\emlineto{55.664}{40.914} 
\emmoveto{55.664}{40.904} 
\emlineto{56.254}{37.687} 
\emlineto{56.254}{37.697} 
\emmoveto{56.254}{37.687} 
\emlineto{56.844}{37.755} 
\emlineto{56.844}{37.765} 
\emmoveto{56.844}{37.755} 
\emlineto{57.434}{38.461} 
\emlineto{57.434}{38.471} 
\emmoveto{57.434}{38.461} 
\emlineto{58.024}{38.565} 
\emlineto{58.024}{38.575} 
\emmoveto{58.024}{38.565} 
\emlineto{58.614}{43.066} 
\emlineto{58.614}{43.076} 
\emmoveto{58.614}{43.066} 
\emlineto{59.204}{43.254} 
\emlineto{59.204}{43.264} 
\emmoveto{59.204}{43.254} 
\emlineto{59.794}{40.108} 
\emlineto{59.794}{40.118} 
\emmoveto{59.794}{40.108} 
\emlineto{60.383}{42.921} 
\emlineto{60.383}{42.931} 
\emmoveto{60.383}{42.921} 
\emlineto{60.973}{39.554} 
\emlineto{60.973}{39.564} 
\emmoveto{60.973}{39.554} 
\emlineto{61.563}{43.066} 
\emlineto{61.563}{43.076} 
\emmoveto{61.563}{43.066} 
\emlineto{62.153}{37.225} 
\emlineto{62.153}{37.235} 
\emmoveto{62.153}{37.225} 
\emlineto{62.743}{37.718} 
\emlineto{62.743}{37.728} 
\emmoveto{62.743}{37.718} 
\emlineto{63.333}{41.984} 
\emlineto{63.333}{41.994} 
\emmoveto{63.333}{41.984} 
\emlineto{63.923}{43.571} 
\emlineto{63.923}{43.581} 
\emmoveto{63.923}{43.571} 
\emlineto{64.513}{43.632} 
\emlineto{64.513}{43.642} 
\emmoveto{64.513}{43.632} 
\emlineto{65.103}{44.058} 
\emlineto{65.103}{44.068} 
\emmoveto{65.103}{44.058} 
\emlineto{65.693}{44.094} 
\emlineto{65.693}{44.104} 
\emmoveto{65.693}{44.094} 
\emlineto{66.283}{46.165} 
\emlineto{66.283}{46.175} 
\emmoveto{66.283}{46.165} 
\emlineto{66.873}{45.523} 
\emlineto{66.873}{45.533} 
\emmoveto{66.873}{45.523} 
\emlineto{67.463}{50.171} 
\emlineto{67.463}{50.181} 
\emmoveto{67.463}{50.171} 
\emlineto{68.053}{49.924} 
\emlineto{68.053}{49.934} 
\emmoveto{68.053}{49.924} 
\emlineto{68.643}{47.588} 
\emlineto{68.643}{47.598} 
\emmoveto{68.643}{47.588} 
\emlineto{69.233}{48.644} 
\emlineto{69.233}{48.654} 
\emmoveto{69.233}{48.644} 
\emlineto{69.823}{43.611} 
\emlineto{69.823}{43.621} 
\emmoveto{69.823}{43.611} 
\emlineto{70.413}{48.048} 
\emlineto{70.413}{48.058} 
\emmoveto{70.413}{48.048} 
\emlineto{71.003}{43.798} 
\emlineto{71.003}{43.808} 
\emmoveto{71.003}{43.798} 
\emlineto{71.593}{42.017} 
\emlineto{71.593}{42.027} 
\emmoveto{71.593}{42.017} 
\emlineto{72.183}{43.194} 
\emlineto{72.183}{43.204} 
\emmoveto{72.183}{43.194} 
\emlineto{72.773}{50.388} 
\emlineto{72.773}{50.398} 
\emmoveto{72.773}{50.388} 
\emlineto{73.363}{48.351} 
\emlineto{73.363}{48.361} 
\emmoveto{73.363}{48.351} 
\emlineto{73.953}{49.308} 
\emlineto{73.953}{49.318} 
\emmoveto{73.953}{49.308} 
\emlineto{74.543}{49.394} 
\emlineto{74.543}{49.404} 
\emmoveto{74.543}{49.394} 
\emlineto{75.133}{50.432} 
\emlineto{75.133}{50.442} 
\emmoveto{75.133}{50.432} 
\emlineto{75.723}{47.048} 
\emlineto{75.723}{47.058} 
\emmoveto{75.723}{47.048} 
\emlineto{76.313}{48.115} 
\emlineto{76.313}{48.125} 
\emmoveto{76.313}{48.115} 
\emlineto{76.903}{55.801} 
\emlineto{76.903}{55.811} 
\emmoveto{76.903}{55.801} 
\emlineto{77.493}{49.130} 
\emlineto{77.493}{49.140} 
\emmoveto{77.493}{49.130} 
\emlineto{78.083}{52.829} 
\emlineto{78.083}{52.839} 
\emmoveto{78.083}{52.829} 
\emlineto{78.673}{45.917} 
\emlineto{78.673}{45.927} 
\emmoveto{78.673}{45.917} 
\emlineto{79.263}{51.335} 
\emlineto{79.263}{51.345} 
\emmoveto{79.263}{51.335} 
\emlineto{79.853}{54.851} 
\emlineto{79.853}{54.861} 
\emmoveto{79.853}{54.851} 
\emlineto{80.442}{45.212} 
\emlineto{80.442}{45.222} 
\emmoveto{80.442}{45.212} 
\emlineto{81.032}{47.958} 
\emlineto{81.032}{47.968} 
\emmoveto{81.032}{47.958} 
\emlineto{81.622}{51.082} 
\emlineto{81.622}{51.092} 
\emmoveto{81.622}{51.082} 
\emlineto{82.212}{53.685} 
\emlineto{82.212}{53.695} 
\emmoveto{82.212}{53.685} 
\emlineto{82.802}{54.851} 
\emlineto{82.802}{54.861} 
\emmoveto{82.802}{54.851} 
\emlineto{83.392}{52.713} 
\emlineto{83.392}{52.723} 
\emmoveto{83.392}{52.713} 
\emlineto{83.982}{54.600} 
\emlineto{83.982}{54.610} 
\emmoveto{83.982}{54.600} 
\emlineto{84.572}{51.972} 
\emlineto{84.572}{51.982} 
\emmoveto{84.572}{51.972} 
\emlineto{85.162}{53.169} 
\emlineto{85.162}{53.179} 
\emmoveto{85.162}{53.169} 
\emlineto{85.752}{58.328} 
\emlineto{85.752}{58.338} 
\emmoveto{85.752}{58.328} 
\emlineto{86.342}{57.245} 
\emlineto{86.342}{57.255} 
\emmoveto{86.342}{57.245} 
\emlineto{86.932}{56.568} 
\emlineto{86.932}{56.578} 
\emmoveto{86.932}{56.568} 
\emlineto{87.522}{55.182} 
\emlineto{87.522}{55.192} 
\emmoveto{87.522}{55.182} 
\emlineto{88.112}{55.153} 
\emlineto{88.112}{55.163} 
\emmoveto{88.112}{55.153} 
\emlineto{88.702}{58.290} 
\emlineto{88.702}{58.300} 
\emmoveto{88.702}{58.290} 
\emlineto{89.292}{51.637} 
\emlineto{89.292}{51.647} 
\emmoveto{89.292}{51.637} 
\emlineto{89.882}{55.024} 
\emlineto{89.882}{55.034} 
\emmoveto{89.882}{55.024} 
\emlineto{90.472}{54.306} 
\emlineto{90.472}{54.316} 
\emmoveto{90.472}{54.306} 
\emlineto{91.062}{56.142} 
\emlineto{91.062}{56.152} 
\emmoveto{91.062}{56.142} 
\emlineto{91.652}{59.613} 
\emlineto{91.652}{59.623} 
\emmoveto{91.652}{59.613} 
\emlineto{92.242}{58.027} 
\emlineto{92.242}{58.037} 
\emmoveto{92.242}{58.027} 
\emlineto{92.832}{59.542} 
\emlineto{92.832}{59.552} 
\emmoveto{92.832}{59.542} 
\emlineto{93.422}{59.396} 
\emlineto{93.422}{59.406} 
\emmoveto{93.422}{59.396} 
\emlineto{94.012}{57.481} 
\emlineto{94.012}{57.491} 
\emmoveto{94.012}{57.481} 
\emlineto{94.602}{65.182} 
\emlineto{94.602}{65.192} 
\emmoveto{94.602}{65.182} 
\emlineto{95.192}{63.996} 
\emlineto{95.192}{64.006} 
\emmoveto{95.192}{63.996} 
\emlineto{95.782}{58.926} 
\emlineto{95.782}{58.936} 
\emmoveto{95.782}{58.926} 
\emlineto{96.372}{60.813} 
\emlineto{96.372}{60.823} 
\emmoveto{96.372}{60.813} 
\emlineto{96.962}{57.376} 
\emlineto{96.962}{57.386} 
\emmoveto{96.962}{57.376} 
\emlineto{97.552}{62.411} 
\emlineto{97.552}{62.421} 
\emmoveto{97.552}{62.411} 
\emlineto{98.142}{57.859} 
\emlineto{98.142}{57.869} 
\emmoveto{98.142}{57.859} 
\emlineto{98.732}{57.498} 
\emlineto{98.732}{57.508} 
\emmoveto{98.732}{57.498} 
\emlineto{99.322}{58.568} 
\emlineto{99.322}{58.578} 
\emmoveto{99.322}{58.568} 
\emlineto{99.912}{64.196} 
\emlineto{99.912}{64.206} 
\emmoveto{99.912}{64.196} 
\emlineto{100.501}{63.914} 
\emlineto{100.501}{63.924} 
\emmoveto{100.501}{63.914} 
\emlineto{101.091}{62.508} 
\emlineto{101.091}{62.518} 
\emmoveto{101.091}{62.508} 
\emlineto{101.681}{61.306} 
\emlineto{101.681}{61.316} 
\emmoveto{101.681}{61.306} 
\emlineto{102.271}{66.329} 
\emlineto{102.271}{66.339} 
\emmoveto{102.271}{66.329} 
\emlineto{102.861}{63.500} 
\emlineto{102.861}{63.510} 
\emmoveto{102.861}{63.500} 
\emlineto{103.451}{65.234} 
\emlineto{103.451}{65.244} 
\emmoveto{103.451}{65.234} 
\emlineto{104.041}{70.951} 
\emlineto{104.041}{70.961} 
\emmoveto{104.041}{70.951} 
\emlineto{104.631}{64.204} 
\emlineto{104.631}{64.214} 
\emmoveto{104.631}{64.204} 
\emlineto{105.221}{68.282} 
\emlineto{105.221}{68.292} 
\emmoveto{105.221}{68.282} 
\emlineto{105.811}{61.678} 
\emlineto{105.811}{61.688} 
\emmoveto{105.811}{61.678} 
\emlineto{106.401}{65.698} 
\emlineto{106.401}{65.708} 
\emmoveto{106.401}{65.698} 
\emlineto{106.991}{67.541} 
\emlineto{106.991}{67.551} 
\emmoveto{106.991}{67.541} 
\emlineto{107.581}{61.428} 
\emlineto{107.581}{61.438} 
\emmoveto{107.581}{61.428} 
\emlineto{108.171}{60.937} 
\emlineto{108.171}{60.947} 
\emmoveto{108.171}{60.937} 
\emlineto{108.761}{68.715} 
\emlineto{108.761}{68.725} 
\emmoveto{108.761}{68.715} 
\emlineto{109.351}{68.013} 
\emlineto{109.351}{68.023} 
\emmoveto{109.351}{68.013} 
\emlineto{109.941}{70.822} 
\emlineto{109.941}{70.832} 
\emmoveto{109.941}{70.822} 
\emlineto{110.531}{67.367} 
\emlineto{110.531}{67.377} 
\emmoveto{110.531}{67.367} 
\emlineto{111.121}{70.504} 
\emlineto{111.121}{70.514} 
\emmoveto{111.121}{70.504} 
\emlineto{111.711}{66.945} 
\emlineto{111.711}{66.955} 
\emmoveto{111.711}{66.945} 
\emlineto{112.301}{68.940} 
\emlineto{112.301}{68.950} 
\emmoveto{112.301}{68.940} 
\emlineto{112.891}{75.344} 
\emlineto{112.891}{75.354} 
\emmoveto{112.891}{75.344} 
\emlineto{113.481}{70.812} 
\emlineto{113.481}{70.822} 
\emmoveto{113.481}{70.812} 
\emlineto{114.071}{73.616} 
\emlineto{114.071}{73.626} 
\emmoveto{114.071}{73.616} 
\emlineto{114.661}{67.989} 
\emlineto{114.661}{67.999} 
\emmoveto{114.661}{67.989} 
\emlineto{115.251}{70.226} 
\emlineto{115.251}{70.236} 
\emmoveto{115.251}{70.226} 
\emlineto{115.841}{73.933} 
\emlineto{115.841}{73.943} 
\emmoveto{115.841}{73.933} 
\emlineto{116.431}{65.408} 
\emlineto{116.431}{65.418} 
\emmoveto{116.431}{65.408} 
\emlineto{117.021}{69.867} 
\emlineto{117.021}{69.877} 
\emmoveto{117.021}{69.867} 
\emlineto{117.611}{67.433} 
\emlineto{117.611}{67.443} 
\emmoveto{117.611}{67.433} 
\emlineto{118.201}{71.141} 
\emlineto{118.201}{71.151} 
\emmoveto{118.201}{71.141} 
\emlineto{118.791}{74.154} 
\emlineto{118.791}{74.164} 
\emmoveto{118.791}{74.154} 
\emlineto{119.381}{71.352} 
\emlineto{119.381}{71.362} 
\emmoveto{119.381}{71.352} 
\emlineto{119.971}{69.853} 
\emlineto{119.971}{69.863} 
\emmoveto{119.971}{69.853} 
\emlineto{120.560}{72.316} 
\emlineto{120.560}{72.326} 
\emmoveto{120.560}{72.316} 
\emlineto{121.150}{70.659} 
\emlineto{121.150}{70.669} 
\emmoveto{121.150}{70.659} 
\emlineto{121.740}{77.910} 
\emlineto{121.740}{77.920} 
\emmoveto{121.740}{77.910} 
\emlineto{122.330}{76.822} 
\emlineto{122.330}{76.832} 
\emmoveto{122.330}{76.822} 
\emlineto{122.920}{74.160} 
\emlineto{122.920}{74.170} 
\emmoveto{122.920}{74.160} 
\emlineto{123.510}{76.373} 
\emlineto{123.510}{76.383} 
\emmoveto{123.510}{76.373} 
\emlineto{124.100}{73.699} 
\emlineto{124.100}{73.709} 
\emmoveto{124.100}{73.699} 
\emlineto{124.690}{76.928} 
\emlineto{124.690}{76.938} 
\emmoveto{124.690}{76.928} 
\emlineto{125.280}{71.586} 
\emlineto{125.280}{71.596} 
\emmoveto{125.280}{71.586} 
\emlineto{125.870}{72.180} 
\emlineto{125.870}{72.190} 
\emmoveto{125.870}{72.180} 
\emlineto{126.460}{76.340} 
\emlineto{126.460}{76.350} 
\emmoveto{126.460}{76.340} 
\emlineto{127.050}{76.714} 
\emlineto{127.050}{76.724} 
\emmoveto{127.050}{76.714} 
\emlineto{127.640}{76.640} 
\emlineto{127.640}{76.650} 
\emmoveto{127.640}{76.640} 
\emlineto{128.230}{77.077} 
\emlineto{128.230}{77.087} 
\emmoveto{128.230}{77.077} 
\emlineto{128.820}{77.805} 
\emlineto{128.820}{77.815} 
\emmoveto{128.820}{77.805} 
\emlineto{129.410}{79.988} 
\emlineto{129.410}{79.998} 
\emshow{24.980}{25.400}{} 
\emshow{1.000}{10.000}{9.97e-2} 
\emshow{1.000}{24.000}{1.01e-1} 
\emshow{1.000}{38.000}{1.02e-1} 
\emshow{1.000}{52.000}{1.03e-1} 
\emshow{1.000}{66.000}{1.04e-1} 
\emshow{1.000}{80.000}{1.05e-1} 
\emshow{12.000}{5.000}{0.00e0} 
\emshow{23.800}{5.000}{1.00e0} 
\emshow{35.600}{5.000}{2.00e0} 
\emshow{47.400}{5.000}{3.00e0} 
\emshow{59.200}{5.000}{4.00e0} 
\emshow{71.000}{5.000}{5.00e0} 
\emshow{82.800}{5.000}{6.00e0} 
\emshow{94.600}{5.000}{7.00e0} 
\emshow{106.400}{5.000}{8.00e0} 
\emshow{118.200}{5.000}{9.00e0} 
\emshow{130.000}{5.000}{1.00e1}

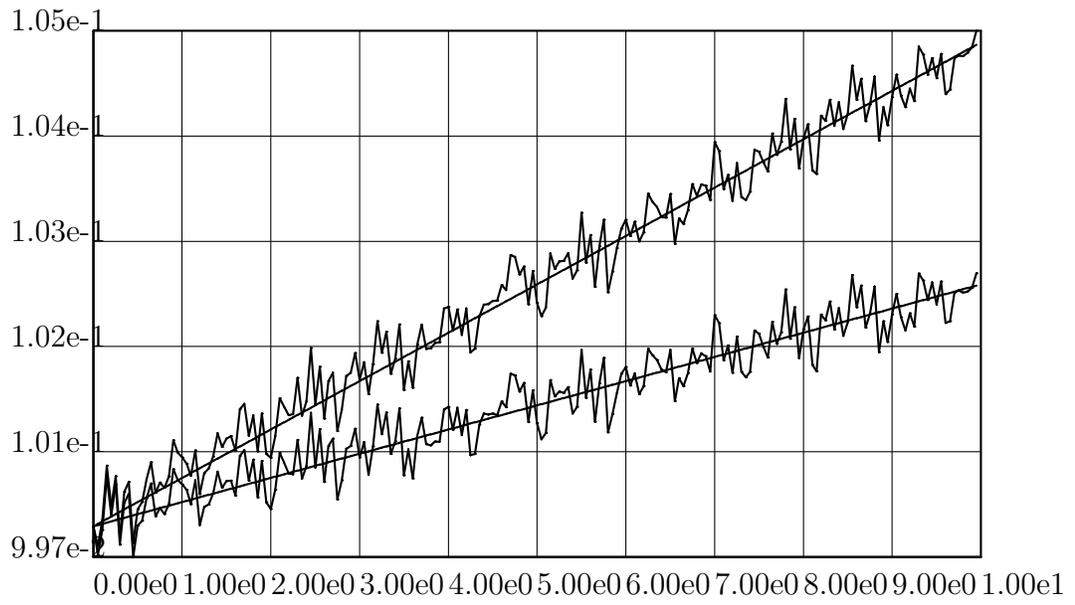
{\centerline {\bf Fig. 1}}

 \end{document}